\documentclass[twocolumn,aps,pre,showpacs]{revtex4}
\usepackage{amsmath}
\usepackage{amsfonts}
\usepackage{graphicx}
\graphicspath{{figures/}}
\usepackage{bm}
\usepackage{color}
\definecolor{darkgreen}{rgb}{0,.5,0}
\usepackage{amssymb}
\usepackage{float}
\usepackage[colorlinks=true,citecolor=darkgreen,linkcolor=blue]{hyperref}
\usepackage[T1,T2A]{fontenc}
\usepackage[english]{babel}
\usepackage{xcolor}

\newcommand{\comment}[1]{}
\newcommand{\rG}{\rm{G}}
\newcommand{\rN}{\rm{N}}

\begin{document}

\title{Explosive synchronization in multiplex neuron-glial networks}

\author{Tetyana Laptyeva$^1$}
\author{Sarika Jalan$^{2}$}
\author{Mikhail Ivanchenko$^3$}

\affiliation{$^1$Department of Control Theory and Systems Dynamics,Lobachevsky State University of
Nizhny Novgorod, Nizhny Novgorod, Russia}
\affiliation{$^2$Complex Systems Lab, Department of Physics, Indian Institute of Technology Indore, Simrol, Indore-452020}
\affiliation{$^3$Department of Applied Mathematics and Laboratory of Systems Medicine of Healthy Aging, Lobachevsky State University of Nizhny Novgorod, Nizhny Novgorod, Russia}

\begin {abstract}
Explosive synchronization refers to an abrupt (first order) transition to non-zero phase order parameter in oscillatory networks, underpinned by the bistability of synchronous and asynchronous states. Growing evidence suggests that this phenomenon might be no less general then the celebrated Kuramoto scenario that belongs to the second order universality class. Importantly, the recent examples demonstrate that explosive synchronization can occur for certain network topologies and coupling types, like the global higher-order coupling, without specific requirements on the individial oscillator dynamics or dynamics-network correlations. Here we demonstrate a rich picture of explosive synchronization and desynchronization transitions in multiplex networks, where it is sufficient to have a single random sparsly connected layer with higher-order coupling terms (and not necessarily in the synchronization regime on its own), the other layer being a regular lattice without own phase transitions at all. Moreover, explosive synchronization emerges even when the random layer has only low-order pairwise coupling, althoug the hysteresis interval becomes narrow and explosive desynchronization is no longer observed. The relevance to the normal and pathological dynamics of neural-glial networks is pointed out.     
\end{abstract}


\maketitle

\section{Introduction}

Continuous (``soft'') and discontinous (``hard'') transitions to synchronization have stayed in the focus of attention since the early studies of the phenomenon \cite{Appleton1922,vdP1927,Andronov1930a,Andronov1930b}. There the onset of zero frequency mismatch is underpinned by different bifurcation scenarious, the main control parameters being the driving amplitude or mutual coupling strength. Later investigations elaborated the picture regarding bi- and multi- stability as a source of the hard scenario \cite{Rand1980,Aronson1990,Ivanchenko2004}.

In oscillatory networks, synchronization used to be associated with the Kuramoto-type second-order phase transition, characterized by the continuous increase of the appropriately introduced phase order parameter above the bifurcation point \cite{Kuramoto1975,Strogatz2000}. Until recently, few exceptions permitting an abrupt transition have been known, to name inertia in indivudual phase dynamics \cite{Tanaka1997} and specific frequency distributions \cite{Hemmen1993,Basnarkov2007}. 

The interest in complex networks dynamics opened a new perspective, and, along with generalization of conditions for the second order transition to synchrony \cite{Hong2002,Um2014}, there emerged multiple examples of the first order (``explosive'' or ``abrupt'') transition, underpinned by correlations between  natural frequencies and node degrees of the oscillators
\cite{Moreno2011}, weighting coupling according to the mismatch in natural frequencies of neighboring oscillators, implementing other disassortativity rules \cite{Leyva2013a,Leyva2013b,Zhu2013,Zhang2014}, or introducing adaptive coupling according to local or global phase order parameters \cite{Zhang2015,Dai2020,Kumar2020}.

Despite the progress, there still remained a question, whether explosive synchronization can occur without a specific relation between the network architecture and static or dynamical properties of the nodes. One of the found possibilities was implementing higher order coupling (triplet and quadruplet interactions, or 2- and 3-simpleces, respectively) on top of the all-to-all pairwise coupling \cite{Scardal2020}. Another one was to introduce multiplex networks with attractive and repulsive coupling in different layers, motivated by the physiological reality of biological neural networks \cite{Jalan2019}.    

Multiplex networks have provided a rich framework to observe abrupt sycnrhonization transitions. Beside the above example, it was also demonstrated, that multiplexing layers with the different dynamical properties, inertial and non-inertial Kuramoto oscillations (supporting and non-supporting explosive syncrhonization) can lead to the pan-network first order transition \cite{Jalan2020}. Multiplexing layers with pairwise and higher-order all-to-all intralayer coupling can establish the global explosive synchronization transition as well \cite{Jalan2023}.

\begin{figure*}
	\begin{minipage}[h]{0.49\linewidth}
		\center{\includegraphics[width=0.99\linewidth]{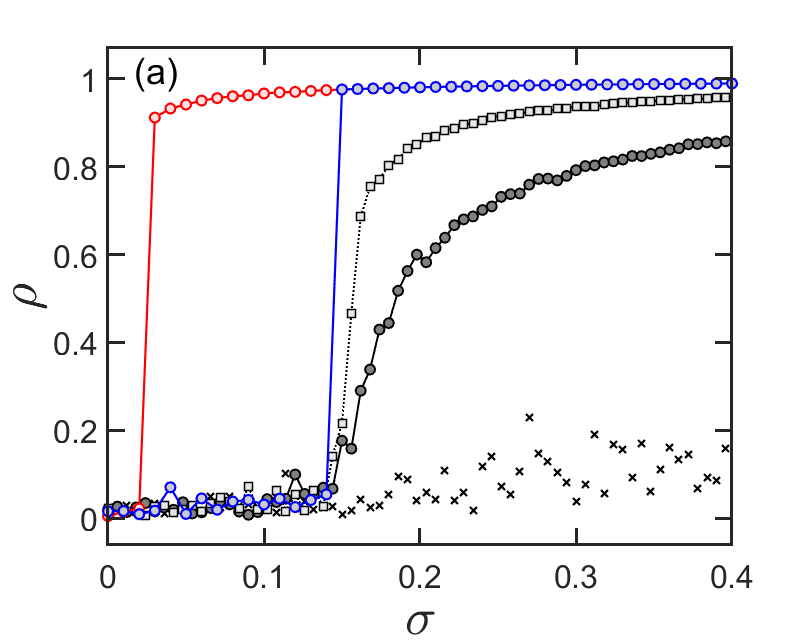}}
	\end{minipage}
	\hfill
	\begin{minipage}[h]{0.49\linewidth}
		\center{\includegraphics[width=\linewidth]{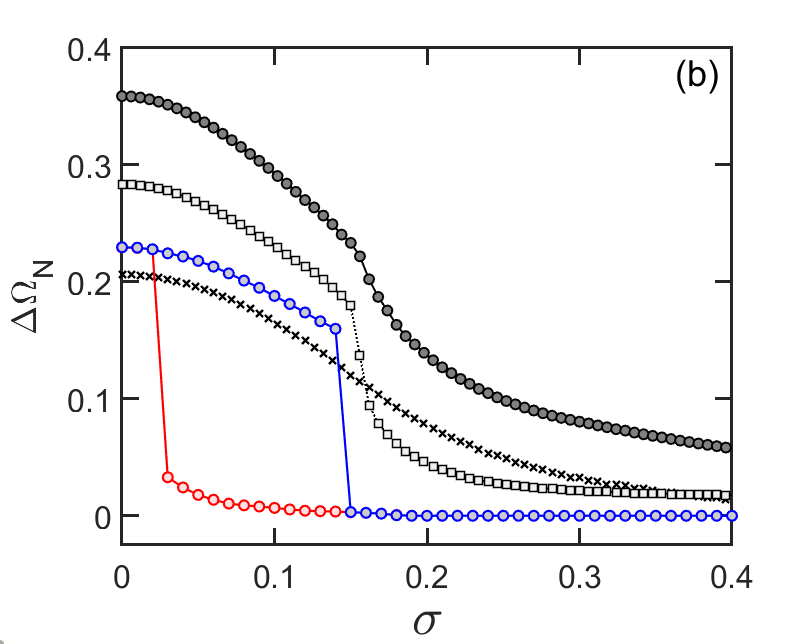}}
	\end{minipage}
	\caption{Kuramoto mean-field (a) and frequency (b) synchronization in the layers of different topology in dependence on the intralayer coupling $\sigma = \sigma_{\rG} = K^{(\rN)}_1$, when the interlayer coupling is absent, $\sigma_{\rG\rN} = 0$. Here <<$\times$>> refer to the regular lattice (glial subsystem), and the other markers to the neural network with triadic interaction strength $K^{(\rN)}_2 = 0.0$ (black <<$\circ$>>), $K^{(\rN)}_2 = 0.05$ (grey <<$\square$>>) and $K^{(\rN)}_2 = 0.15$ (blue and red <<$\circ$>> for increasing and decreasing $\sigma=K^{(\rN)}_1$ passages, respectively). Here the system size is $N = 50 \times 50$.}
	\label{fig:1}
\end{figure*}

In this paper we extend these findings, investigating the two-layer multiplex network that models the dynamics of neural-glial (neural-astrocytic) enembles. The ``neural'' layer can be implemented as a random Erd\"{o}s-R\'{e}nyi graph, or a small-world graph, or an intermediate achitecture \cite{Barabasi}. The ``glial'' layer is implemented as a locally coupled regular lattice, reflecting diffusion of extracellular calcium, glutamate and other mediators of astrocytic interactions. Distributions of the natural frequencies between the layers differ, such that the ``neural'' oscillators frequencies are an order of magnitude greater than that of ``astrocytic''. Previoulsy, it has been demonstrated, that beside a quite intuitive transduction of the second order transition to the regular lattice layer (where it is impossible on its own), the interaction between the layers can induce such a transition even if the ``neural'' layer is not originally synchronized \cite{Ma2017,Makovkin2021}. 

First, we address the case of the higher-order and irregular sparse coupling in the ``neural'' layer and investigate if multiplexing to a ``glial'' lattice can induce global explosive transition. Second, we demonstrate that even if the ``neural'' layer has just pairwise interactions (that permits the second order transition only),  multiplexing it to the lattice (that does not allow for the phase order at all) can initiate global explosive-type syncrhony under certain conditions.

\section{Model}
We study the two-layer network  model of phase oscillators, one of which mimics an ensemble of interacting glial cells (astrocytes) and the other layer represents interacting neural cells. This is a qualitative model of the cerebral cortex, that captures the multi-timescale nature and multiplex topology of biological neuronal and glial networks\cite{Ma2017,Makovkin2021}. Both layers consist of $N = M \times M$ nodes. 

Interactions between glial and glial-neural cells are local since in the physiological systems they are provided by diffusion of extracellular calcium, glutamate and other neurotransmitters. We assume that ``glial'' oscillators are placed on a two-dimensional quadratic lattice.

\begin{figure*}
	\begin{minipage}[h]{0.49\linewidth}
		\center{\includegraphics[width=\linewidth]{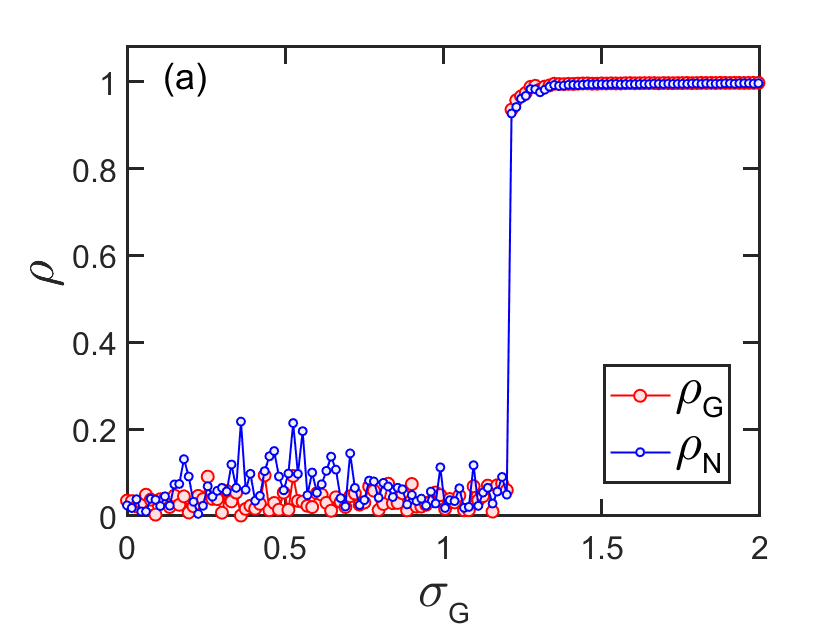}}
	\end{minipage}
	\begin{minipage}[h]{0.49\linewidth}
		\center{\includegraphics[width=\linewidth]{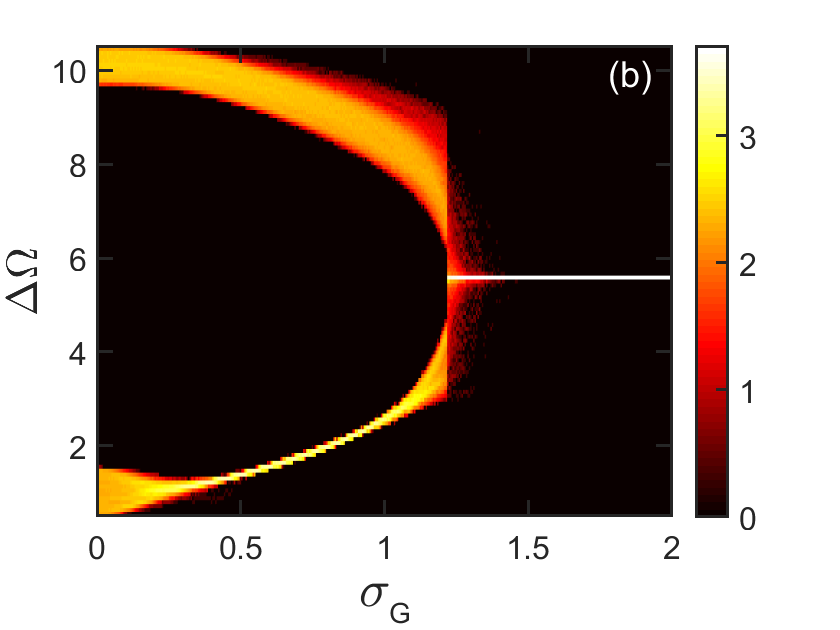}}
	\end{minipage}
	\caption{Mean field (a) and color-coded log10 density distribution of observed frequencies (b) vs. glial coupling strength, $\sigma_{\rG\rN} = \sigma_{\rG}$ in the layers of different topology for the system size $N = 50 \times 50$. Here interactions in the neural network are pairwise and triadic with the coupling strengths $K^{(\rN)}_1 = 0.1$ and $K^{(\rN)}_2 = 0.15$, respectively. At $K^{(\rN)}_1 = 0.1$ oscillations in the isolated neural layer are nonsynchronous.}
	\label{fig:2}
\end{figure*}

\begin{figure*}
	\begin{minipage}[h]{0.49\linewidth}
		\center{\includegraphics[width=\linewidth]{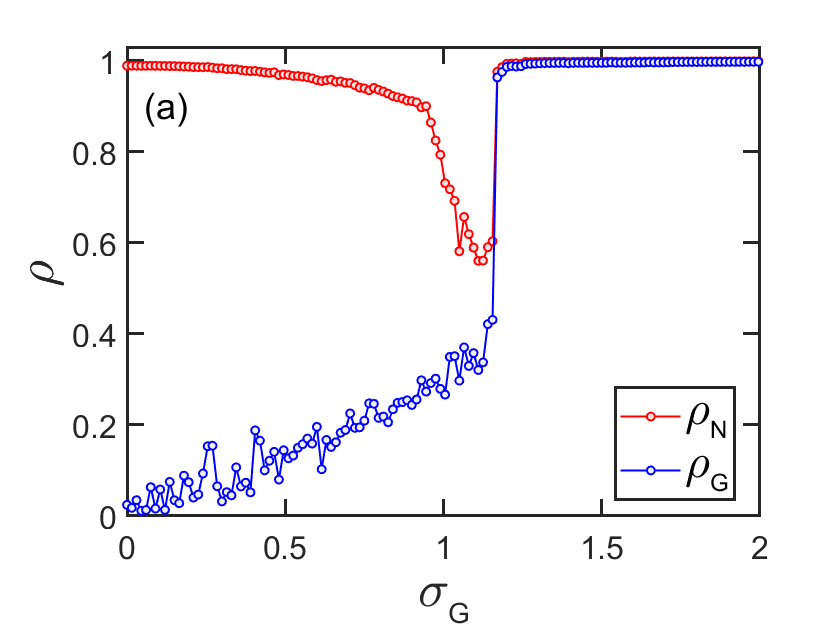}}
	\end{minipage}
	\begin{minipage}[h]{0.49\linewidth}
		\center{\includegraphics[width=\linewidth]{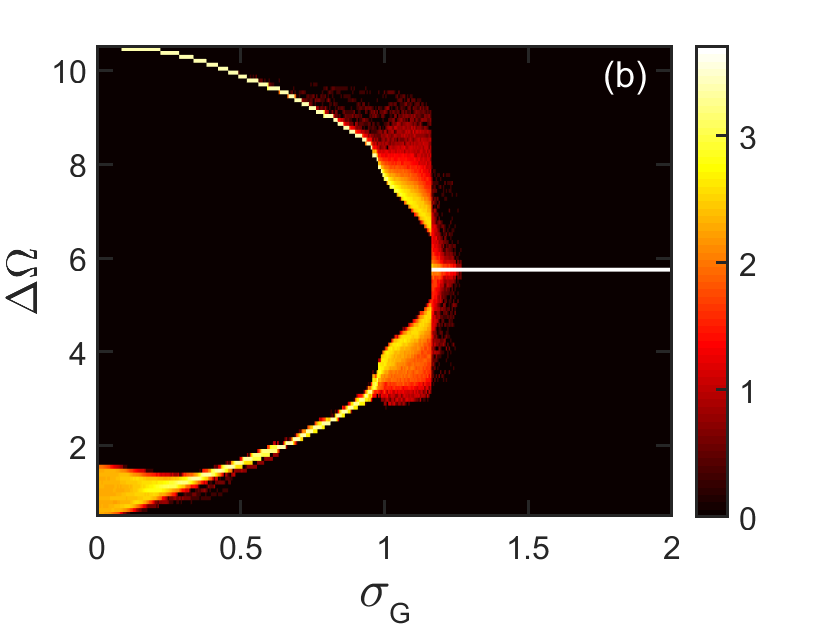}}
	\end{minipage}
	\caption{Mean field (a) and color-coded log10 density distribution of observed frequencies (b) vs. glial coupling strength, $\sigma_{\rG\rN} = \sigma_{\rG}$ in the layers of different topology for the system size $N = 50 \times 50$. Here interactions in the neural network are pairwise and triadic with the coupling strengths $K^{(\rN)}_1 = 0.4$ and $K^{(\rN)}_2 = 0.15$, respectively. At $K^{(\rN)}_1 = 0.4$ oscillations in the isolated neural layer are synchronous.}
	\label{fig:3}
\end{figure*}

Neural cells are characterized by long-range synaptic interactions, and there is evidence that a characteristic architecture is on the border between Erd{\H o}s-R{\'e}nyi random network and the Watts-Strogatz small-world network, and the connectivity is sparse \cite{Muller2014}. Besides, experimental evidence suggest that neurons and brain areas are grouped in cliques and can be effectively described as simplical complexes \cite{Giusti2016,Reimann2017,Sizemore2018}. 

It leads us to the following architecture of the sparse ``neural'' layer: a pair of two nodes connected with each other with probability $p = 4/(N-1)$ (basic pairwise interactions, also referred to as a simplicial complex of order 1); similarly, a set of three nodes connected with each other with probability $p = 16/(N-1)^2$ (triadic interactions, referred to as a simplicial complex of order 2). Finally, interlayer links are such that each node in the neural layer is connected to its mirror and all the neighbors of the mirror node in the glial layer.

The coupled Kuramoto oscillators on multilayer networks with each layer presented by a simplicial network follow the model proposed in \cite{Scardal2020}:
\begin{equation}
\frac{d \theta^{(\rN)}_i}{d t} = \omega^{(\rN)}_i + \mathcal{S}^{(\rN)}_i + \sigma_{\rG\rN} \sum\limits_{\left\{i'\right\}} \sin(\theta^{(\rG)}_{i'} - \theta^{(\rN)}_i),
\label{eq:1}
\end{equation}
for neural layer, where 
\begin{gather}
\label{eq:2}
\mathcal{S}^{(\rN)}_i  = {K}^{(\rN)}_1 \sum\limits^{N}_{j=1} A_{ij} \sin\left(\theta^{(\rN)}_j - \theta^{(\rN)}_i\right) \; + \\ \nonumber
+ \frac{{K}^{(\rN)}_2}{2} \sum\limits^{N}_{j,k=1} B_{ijk} \sin \left( 2\theta^{(\rN)}_j - \theta^{(\rN)}_k - \theta^{(\rN)}_i \right). 
\end{gather}

For glial layer equations of motion are given by
\begin{gather}
\frac{d \theta^{(\rG)}_i}{d t} = \omega^{(\rG)}_i \; + \; \sigma_{\rG} \sum\limits_{\left\{n\right\}} \sin(\theta^{(\rG)}_{n} - \theta^{(\rG)}_i) \; + \nonumber \\ 
 + \; \sigma_{\rG\rN} \sum\limits_{\left\{i'\right\}} \sin(\theta^{(\rN)}_{i'} - \theta^{(\rG)}_i). \label{eq:3}
\end{gather}

In equations (\ref{eq:1})--(\ref{eq:3}) $\theta^{(\alpha)}_i$ and $\omega^{(\alpha)}_i$ are, respectively, the phases and natural frequencies of the $i$-th oscillator in the $\alpha = \left\{\rG, \rN\right\}$ layer. The frequencies are taken randomly from the uniform distribution so that $\omega_i \in [\omega_0^{(\rG,\rN)}-1/2; \omega_0^{(\rG,\rN)}+1/2]$. The neuronal spiking is up to ten times faster than the chemical dynamics of glial cells, therefore we set the mean frequencies $\omega_0^{(\rG)} = 1$ and $\omega_0^{(\rN)} = 10$. $K^{(\rN)}_i$ indicates the coupling strength for the $i$-simplex interactions in the neural layer. $A_{ij}, B_{ijk}$ are adjacency matrix elements $\{0,1\}$ and represent, respectively, pair-wise and triadic interactions. The coupling strengths takes specific values, $\sigma_{\rG}$ and $\sigma_{\rG\rN}$, for interglial and glial-neural interactions, respectively. Finally, we make use of open boundary conditions for both layers.

\begin{figure*}
	\center{\includegraphics[width=1.01\linewidth]{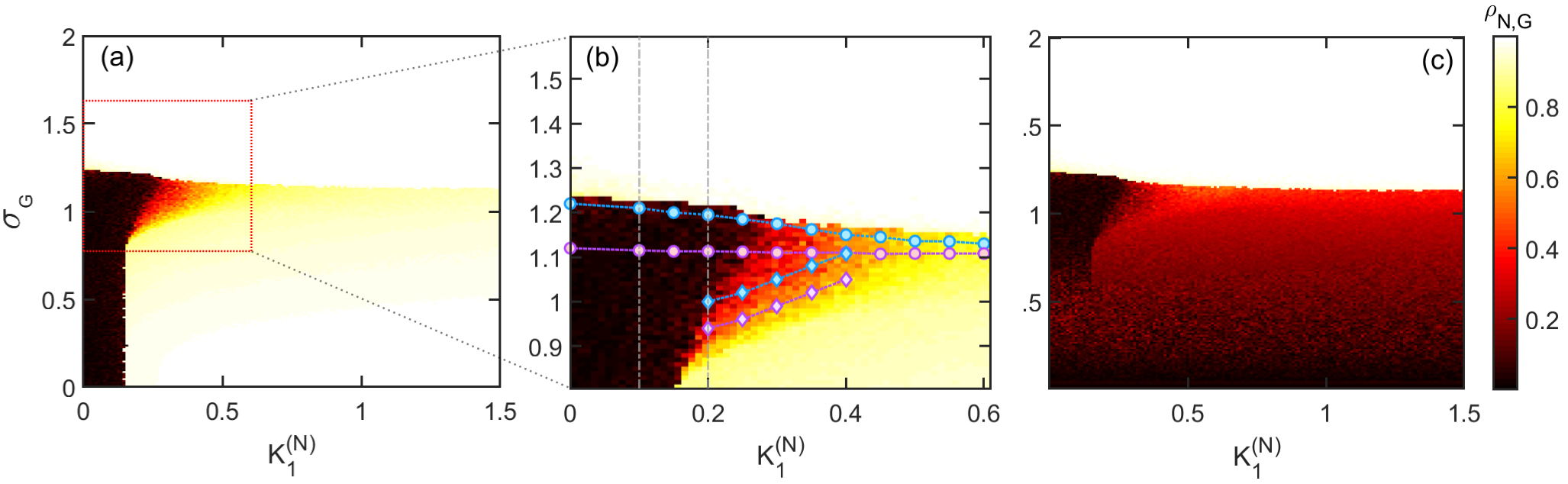}}
	\caption{Two-parameter diagrams for the color-coded subnetwork order parameters, $\rho_{\rN}$ (a) and $\rho_{\rG}$ (c); (b) a zoomed region of two-parameter diagram (a): vertical dashed lines denote forward and backward passages for $K^{(\rN)}_1=0.1$ and $K^{(\rN)}_1=0.1$, cf. Fig.\ref{fig:5}; lines with diamonds (``$\diamond$'') correspond to desynchronization transitions and lines with circles (``$\circ$'') correspond to synchronization transions on the forward (increasing $\sigma_{\rG}$, blue) and backward (decreasing $\sigma_{\rG}$, red) passages. Here the system size is $N = 50 \times 50$ and interactions are pairwise and triadic with the coupling strengths $K^{(\rN)}_1$ and $K^{(\rN)}_2 = 0.15$, respectively.}
\label{fig:4}
\end{figure*}

We use two measures to quantitatively describe synchronization in the network. The global phase order is measured by the Kuramoto order parameter 
\begin{equation}
    \rho = \frac{1}{N} \left| \sum\limits^{N}_{n=1} {\rm e}^{{\rm i} \theta_n} \right|,
\label{eq:4}
\end{equation}
where $\rho$ varies as $0 \leq \rho \leq 1$. The minimal value $\rho = 0$ would realize if all phases are distributed uniformly over the unit circle (oscillators are completely asynchronous), whereas $\rho = 1$ occurs when all phases are identical (oscillators are completely synchronized). The order parameters for each layer were also investigated: $\rho_{\rG}$ and $\rho_{\rN}$ for the glial and neural networks, respectively.

The averaged observed frequency of some $i$-th oscillator is defined as:
\begin{equation}
\Omega_i = \frac{\theta_i(t) - \theta_i(t_0)} {t - t_0},
\label{eq:5}
\end{equation}
where $t$ is some time such that $t>t_0$, and $t_0$ is taken large enough so that the transients would definitely completed. The degree of frequency synchronization is characterized by the standard deviation of observed frequencies (\ref{eq:5}): $\Delta$ for the whole network, $\Delta_{\rG}$ and $\Delta_{\rN}$ for each separated layer, respectively.

The numerical integration of the system with $M = 50$ over a time interval  $[0,T]$ was implemented with a fourth-order Runge-Kutta scheme with the final time $T = 1500$ t.u., time step $\Delta t = 0.001$ and the transient time $t_0 = 500$ t.u. Initial phases were taken randomly from the uniform distribution on a circle.

\section{Results}

\subsection*{Single layer networks}

When the two layers are isolated ($\sigma_{\rG\rN} = 0$), transitions to synchronization are dictated by their respective topology. 
For the isolated regular glial layer the Kuramoto transition does not occur (Fig.\ref{fig:1}a). In the absence of simplex interactions ($K^{(\rN)}_2 = 0$), the isolated random neural network demonstrates the Kuramoto-type transition to phase order at some critical value $K^{\rm synch}_1 \approx 1.5$ (Fig.\ref{fig:1}a). 
The partial frequency synchronization in the neural and glial layers sets on gradually as the coupling is increased (see Fig.\ref{fig:1}b).

In \cite{Scardal2020}, it was demonstrated, in particular, that the 3-layer multiplex network (random $1-$, $2-$ and $3-$simplex interactions in each layer, respectively) can support explosive synchronization. We attempted to further soften this requirement, reducing it to a single-layer network that combiness random $1-$ and $2-$simplex interactions. Numerical simulations show that the presence of sufficiently strong second-order interactions $K^{(\rN)}_2$ gives rise to abrupt synchronization transition (see Figs.\ref{fig:1}). Markedly, the accompanying frequency sycnhronization is complete, as the variance of average frequency drops down to zero.

\begin{figure}[h]
	\includegraphics[width=0.9\linewidth]{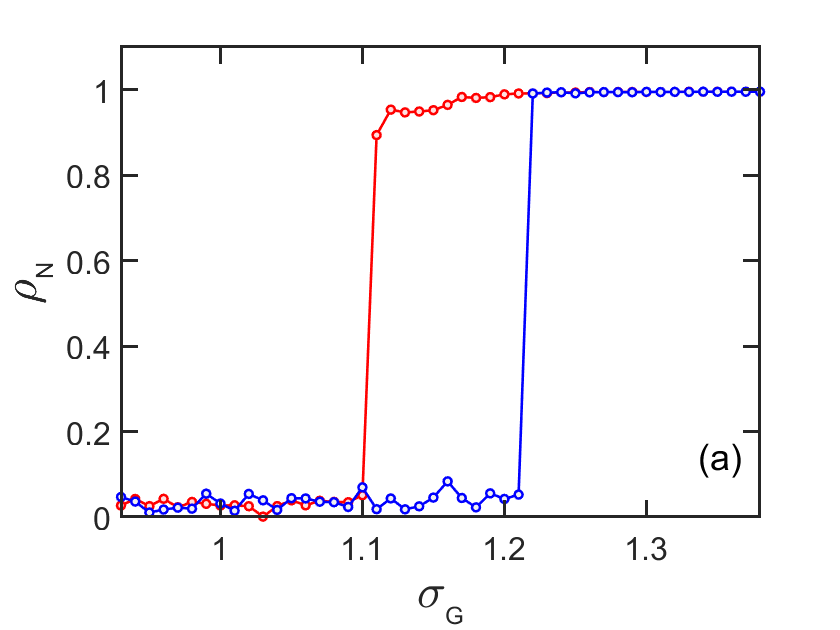} \\
	\includegraphics[width=0.9\linewidth]{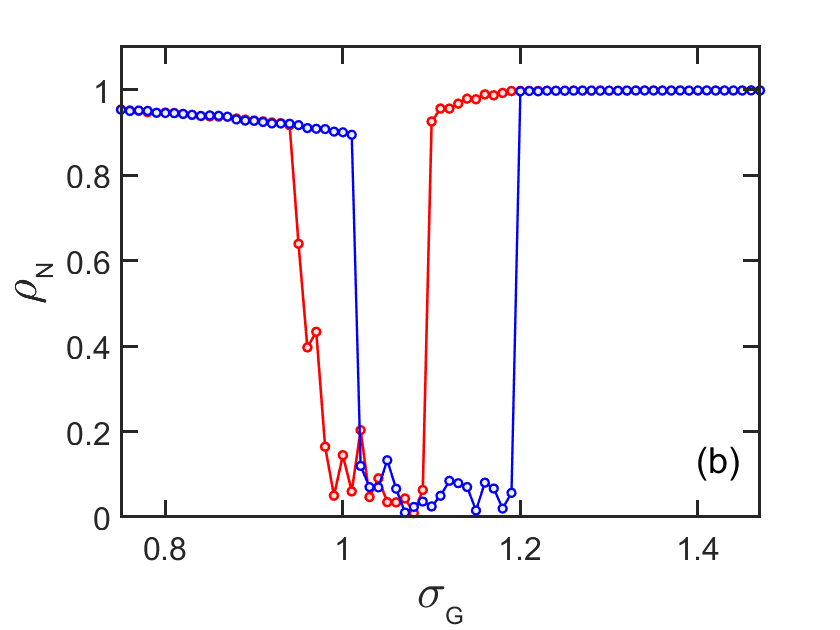}
	\caption{The order parameter $\rho_{\rN}$ is shown as a function of glial coupling strength, $\sigma_{\rG\rN} = \sigma_{\rG}$, in coupled neural and glial layers of size $N = 50 \times 50$. Numerical results are obtained by increasing $\sigma_{\rG}$ forward and subsequent decreasing back, while taking the final set of phases as initial conditions for the next run. This reveals a hysteresis loop with abrupt synchronization and desynchronization transitions. Two typical types of hysteresis loops are shown [passages along the dashed vertical lines in Fig.\ref{fig:4}b]: (a) for $K^{(\rN)}_1 = 0.1$ [see Fig.\ref{fig:2}]; (b) $K^{(\rN)}_1 = 0.2$ [qualitatively similar to the one shown in Fig.\ref{fig:3} for $K^{(\rN)}_1 = 0.4$].}
	\label{fig:5}
\end{figure}

The first-order transition in  underpinned by hysteresis: the system undergoes a transition from a nonsynchronous ($\rho_{\rN} \approx 0.0$) to a synchronous state ($\rho_{\rN} \approx 1.0$) at $K^{\rm synch}_1 \approx 0.15$ as $K^{(\rN)}_1$ is increased (blue line marked by ``o''), and, then, to another abrupt transition from synchronization to desynchronization at $K^{\rm desynch}_1 \approx 0.08$ as $K^{(\rN)}_1$ is decreased (here $K^{(\rN)}_2=0.1$),  (red line marked by ``o''), cf. Fig.\ref{fig:1}.

\subsection*{Multiplexing networks and timescales}

We address the case of multiplex (double layer) networks, coupling the base random (``neural'') layer to the regular (``glial'') one. The latter is additionally characterized by the different (slower) timescale, with regard to the neural-astrocyte physiological reality.

The primary question is, whether multiplexing the layer that supports explosive synchronization (but has not reached synchronization threshold), to the layer that cannot exhibit such transition, can nevertheless induce a global explosive transition. We fix the coupling in the neural layer well below the threshold, $K^{(\rN)}_1 = 0.1$ and $K^{(\rN)}_2 = 0.15$, and gradually increase the interlayer and intra-glial layer coupling strength $\sigma_{\rG\rN} = \sigma_{\rG}$. In result, we observe an abrupt transition at $\sigma_{\rG}\sim1.2$, as evidenced by the order parameter $\rho_{\rN}$ (Fig.\ref{fig:2}(a)). Remarkably, the global order emerges in the regular layer too, $\rho_{\rG}$, and also abruptly. Interestingly, the frequency distribution diagram indicates a non-trivial route to synchronization in the glial layer, where after an almost perfect frequency synchronization, $0.4<\sigma_{\rG}<1.0$, the layer gets desynchronized by the strengthening interaction to the neural layer, ahead of explosive synchronization transition (Fig.\ref{fig:2}(b)). 

Next, we investigate the case when the isolated neural layer is synchronized, 
 $K^{(\rN)}_1 = 0.4$ and $K^{(\rN)}_2 = 0.15$, and the multiplexing coupling $\sigma_{\rG\rN} = \sigma_{\rG}$ is increased. Here we again observe an abrupt transition to synchronization also at about $\sigma_{\rG} \sim 1.2$, preceeded by (i) an almost perfect frequency synchronization with each layer, which, however, remain desynchronized with each other, $0.4<\sigma_{\rG}<1.0$, and (ii) the loss of intralayer synchronization at about $\sigma_{\rG}\sim1.0$ due to interlayer interaction (Fig.\ref{fig:3}). The order parameter $\rho_{\rN}$ noticeably decreases. 

To present a systematics picture, we explore synchronization transitions in the two-parameter space of coupling strengths ($K^{(\rN)}_2$, $\sigma_{\rG} = \sigma_{\rN\rG}$), cf. Fig.\ref{fig:4}. Here, for each parameter value random initial conditions, the frequency and network topology realizations were generated, and the resulting mean field values calculated after a transient period. Beside the previously discussed non-monotonous onset of synchronization as the multiplexing coupling $\sigma_{\rG}=\sigma_{\rG\rN}$ is increased, it is instructive to note the nonzero order parameter emerging in the glial network, $\rho_{\rG}\sim0.4$, for a relatively strong coupling in the neural layer, $K_1^{\rN}\gtrsim0.25$, and under treshold coupling $\sigma_{\rG}<1.0$, Fig.\ref{fig:4}.

The explosive nature of the observed abrupt transition to synchrony is substantiated by the forward and backward passages with varying $\sigma_{\rG}=\sigma_{\rG\rN}$, inherited values of phase variables $\theta_i$, and fixed random frequency and network topology realizations. In result we reveal hysteresis and bistability of synchronized and non-synchronized attractors that underpin explosive synchronization (Fig.\ref{fig:5}). Instructively, for the case when an isolated neural network provides synchronization ($K_1^{(N)}=0.2$, Fig.\ref{fig:5}b), one observes two hysteresis intervals: one related to the onset of global synchronization within $1.1<\sigma_{\rG}<1.2$, and the other related to the desynchronization of the neural layer within $0.95<\sigma_{\rG}<1.02$. Both transitions are explosive. The obtained boundaries are depicted in Fig.\ref{fig:4}b. Note some mismatch between them and the colored regions due to the random choise of frequency and topology realizations for each point in the latter case.   

\begin{figure*}
	\begin{minipage}[h]{0.49\linewidth}
			\includegraphics[width=\linewidth]{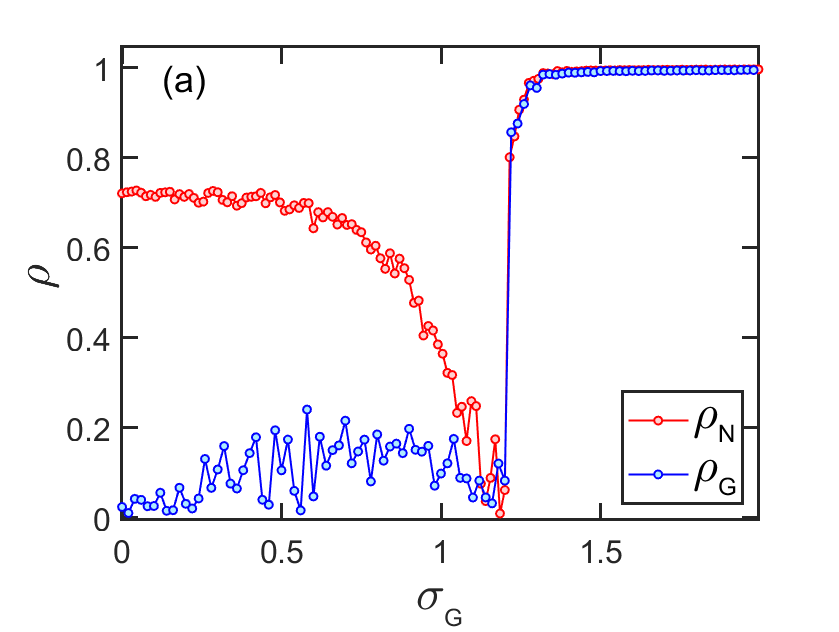}
	\end{minipage}
	\begin{minipage}[h]{0.49\linewidth}
		\includegraphics[width=0.98\linewidth]{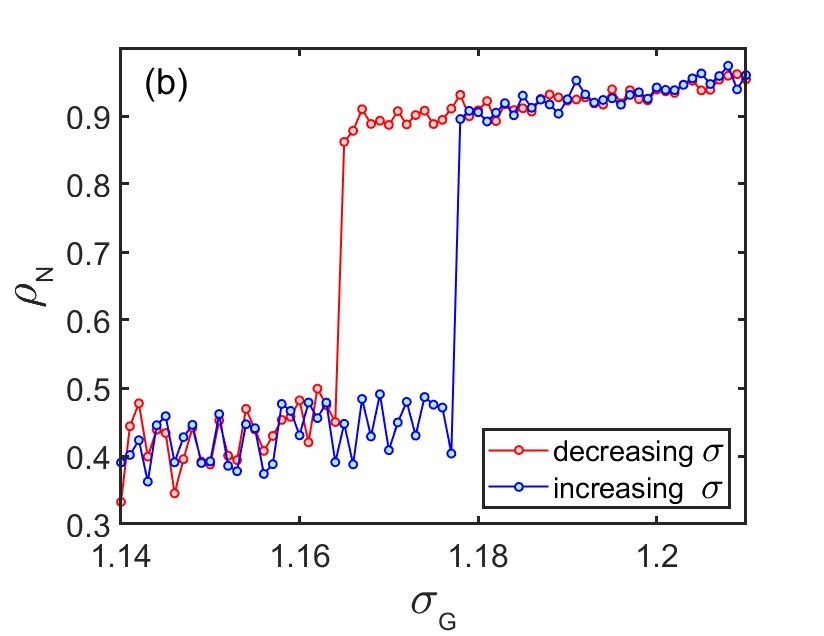}
	\end{minipage}
 
		\begin{minipage}[h]{0.49\linewidth}
		\hspace*{1em}  \includegraphics[width=1.02\linewidth]{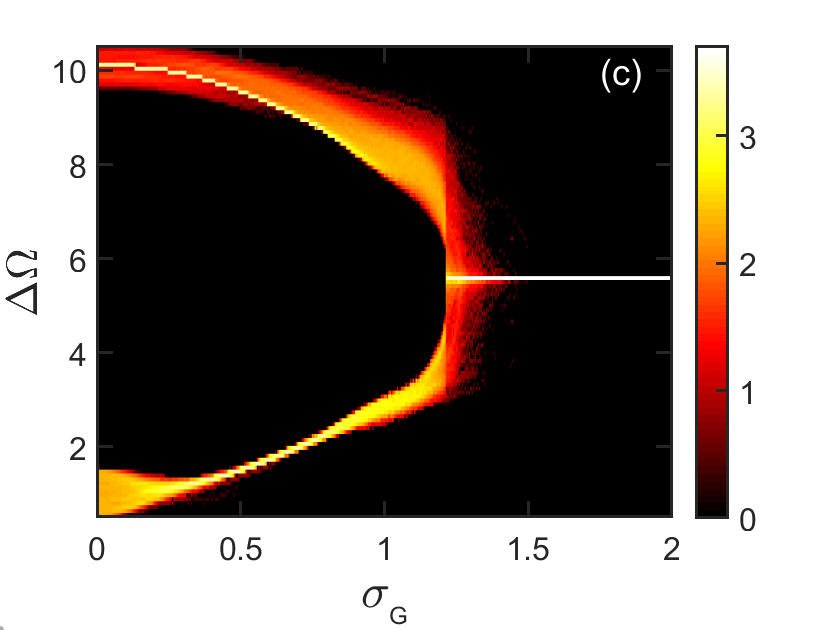}
	\end{minipage}
	\hfill
	\begin{minipage}[h]{0.49\linewidth}
		\hspace*{-1em}\includegraphics[width=1.085\linewidth]{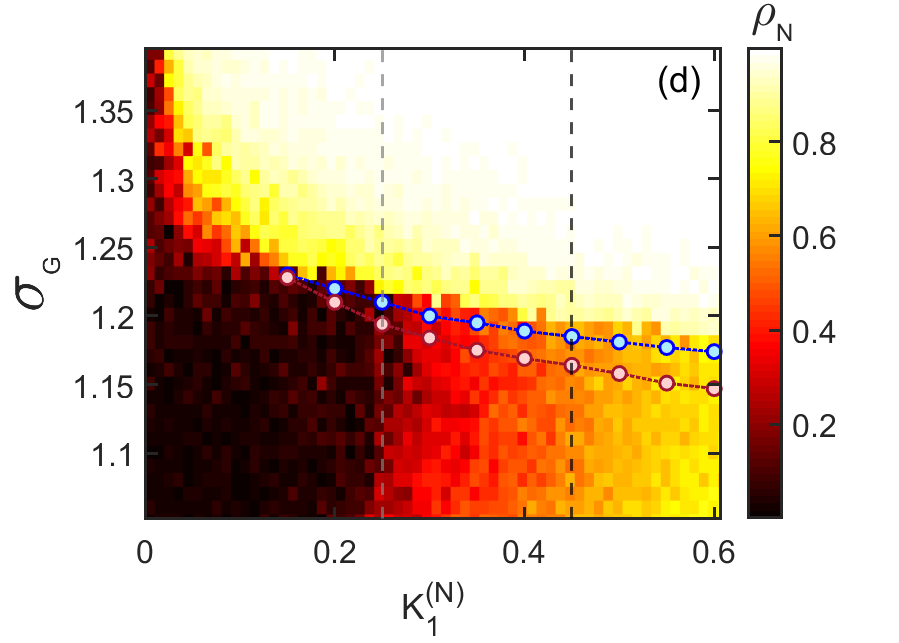}
	\end{minipage}	
	\caption{(a) Mean-field and (c) color-coded log10 density distribution of observed frequencies vs. glial coupling strength, $\sigma_{\rG\rN} = \sigma_{\rG}$, in interacting neural and glial layers for $K^{(\rN)}_1 = 0.25$. (b) The order parameter, $\rho_{\rN}$, is shown as a function of interlayer coupling strength for $K^{(\rN)}_1 = 0.45$ with forward and backward passages over $\sigma_{\rG}\in[1.15,1.22]$ revealing a hysteresis loop and abrupt synchronization and desynchronization transitions at $\sigma_{\rG}^{\rm synch} \approx 1.198$ and $\sigma_{\rG}^{\rm desynch} \approx 0.192$. (d) A zoomed region of two-parameter diagram for the color-coded neural network order parameter, $\rho_{\rN}$. The blue line marked with ``o'' refers to the onset of abrupt synchronization as $\sigma_{\rG}$ increases, and the red line marked with ``o'' shows desynchronization transition along the decrease in $\sigma_{\rG}$. Dashed lines correspond to $K^{(\rN)}_1 = 0.25$ (a) and $K^{(\rN)}_1 = 0.45$ (b). Here neural interactions are pairwise only [$K^{(\rN)}_2 = 0$] and the system size is $N = 50 \times 50$.}
	\label{fig:6}
\end{figure*}

So far, we established the routes to the abrupt synchronization and desynchronization transitions in the random layer with simplical interactions due to multiplex interaction with the glial layer, characterized by the different interlayer coupling topology that does not admit the phase transition to synchrony at all. The difference in frequency scales between the layers has proved to be an essential ingredient, due to desynchronizing effect of the glial layer.

We further hypothesized, that it might be possible to induce an explosive synchronization in this kind of a multiplex network even if none of the isolated layers can support it on its own. In order to verify it, we removed the higher order symplical interactions in the neural layer, setting $K^{(\rN)}_2 = 0.0$, and chose the pairwise interaction strength sufficient to reach a non-zero order parameter in the isolated layer, $K^{(\rN)}_1 = 0.25$. In this setup, the increasing coupling between the internally synchronized neural layer and non-synchronous glial layer can reproduce an abrupt transition to syncrhonization at $\sigma_{\rG}\sim1.2$ after the already familiar intermediate desynchronization interval, $0.7\lesssim\sigma_{\rG}\lesssim1.2$ Fig.\ref{fig:6}(a,c) \cite{Ma2017}.   

Strikingly, a careful forward and backword passiging reveals a hysteresis loop, which occupies in a much smaller parameter range, $1.19\lesssim\sigma_{\rG}\lesssim1.2$, compared to the case of a higher-order simplical network, cf. Fig.\ref{fig:6}b against Fig.\ref{fig:5}, but nevertheless is sufficient to provide an explosive transition. In contrast to the case of a higher-order simplical network, we were not able to observe an explosive desynchronization at $\sigma_{\rG}\approx0.7$. The hysteresis boundaries are shown in Fig.\ref{fig:6}d. Beside a much narrower bistability interval, we also note its collapse at finite  $K^{(\rN)}_1 \approx 0.17$, below which the synchronization transition becomes the second order type.

\section{Conclusion}

Explosive (or abrupt) synchonization in oscillatory networks is associated with the first order phase transition (discontinuous jump in the Kuramoto order parameter), underpinned by the bistability of phase coherent and incoherent regimes. So far it has been reported under a variety of conditions on the dynamics of individual oscillators, often in connection with network properties, to name time delays or node-degree -- oscillator frequency correlations. A more recent work by Arenas and Skardal \cite{Scardal2020} demonstrated that higher-order simplical connectivity in multiplex networks can suffice. Here, guided by the general properties of biological neural-glial networks, we futher investigate the effect of multiplicity, implying that different layers in the multiplex network can be characterized by different timescales (frequency scales) and different topologies (a random network and a regular 2D lattice). Itself, an isolated regular lattice cannot exhibit a global phase coherence, and, in particular, explosive synchronization.

We demonstarte, that multiplexing to such a regular lattice can induce an explosive transition to synchronization in a random higher-order simplectic layer, where an intra-layer coupling is far not sufficient for synchrony. Moreover, a non-zero order parameter emerges in the regular lattice too. Furthermore, multiplexing can break synchronization in the random layer also in an explosive way (with hysteresis), followed by yet another explosive transition back to synchrony, at a greater multiplexing coupling. Ultimately, we find that multiplexing to a regular and frequency detuned lattice can induce a bistability of synchronous and asynchronous states and an explosive transition to synchrony even when the higher-order symplectic coupling is absent, and the random layer contains only pairwise interactions (albeit the width of bistability window is small, and the hysteresis requires a finite intra-layer coupling strength to occur). 

Beside a theoretical advance these results shed futher light on the possible routes to an abrupt onset of synchronization in the real-world systems, in the first place, biological neural networks fucntioning on top of glial cell media, suggesting even less stringent conditions for that, than previously believed. It can be particularty important for understanding and controlling epiliptical seizures, ofter associated with the onset of explosive synchronization.

\section{Acknowledgements}
This work was supported by the Russian Science Foundation Grant No.22-12-00348 (T.L.). Numerical experiments were carried out at the Lobachevsky University supercomputer.


\end{document}